# Determination of the antibiotic susceptibility pattern of Gram-positive bacteria causing UTI in Dhaka, Bangladesh

Bushra Rahman Eipa[1], Md Riadul Islam[1], Raquiba Sultana[1], Seemi Tasnim Alam[1], Tanaj Mehjabin[1], Nohor Noon Haque Bushra[1], S M Moniruzzaman[1], Md Ifrat Hossain[1], Shamia Naz Rashna[1] and Md Aftab Uddin[1*]

*The authors have equal contribution

[1]Department of Microbiology, Stamford University Bangladesh, 51 Siddeswari Road, Dhaka-1217, Bangladesh

*Corresponding author: Dr. Seemi Tasnim Alam, Senior lecturer.
Email: seemitasneem@stamforduniversity.edu.bd

Running title: Antibiotic screening of bacteria causing UTI in Dhaka, Bangladesh

## Abstract

Urinary Tract Infection (UTIs) is referred as one of the most common infection in medical sectors worldwide and antimicrobial resistance (AMR) is also a global threat to human that is related with many diseases. As antibiotics used for the treatment of infectious diseases, the rate of resistance is increasing day by day. Gram positive pathogens are commonly found in urine sample collected from different age groups of people, associated with UTI. The study was conducted in a diagnostic center in Dhaka, Bangladesh with total 1308 urine samples from November 2021 to April 2022. Gram positive pathogens were isolated and antimicrobial susceptibility tests were done. From total 121 samples of gram positive bacteria the highest prevalence rate of UTIs was found in age group of 21-30 year. Mostly *Enterococcus spp.* (33.05%) *Staphylococcus aureus* (27.27%), *Streptococcus* spp. (20.66%), Beta-hemolytic streptococci *(*19.00%) were found as causative agents of UTI compared to others. The majority of isolates have been detected as multi-drug resistant (MDR). The higher percentage of antibiotic resistance were found against Azithromycin (75%), and cefixime (64.46%). This research focused on the regular basis of surveillance for the Gram-positive bacteria antibiotic susceptibility to increase awareness about the use of proper antibiotic thus minimize the drug resistance.

## 1. Introduction

The major cause of urinary tract infection (UTI) is uropathogens (Ullah H *et al*., 2022). Urinary tract infections (UTI) extend their range from uncomplicated cystitis (infection of the bladder) to severe pyelonephritis and nephrolithiasis (kidney stone) and are the third most common types of infection in human medicine worldwide (after respiratory tract infections and infections of the alimentary tract) (Gajdács M *et al*., 2020 & Addis T *et al.*, 2021). Pyelonephritis is the infection of one or both kidneys (Addis T *et al.*, 2021). UTIs also is the second most commonly occurring infections in developed countries, with 100–180 million cases per year (Gajdács M, 2020). Urinary tract infections (UTIs) are very common infectious diseases in the health care system (Gharavi *MJ et al.*, 2021).

The increased morbidity and economic cost caused by UTI affect people of all ages from the neonate to the geriatric age group. Each year, around 150 million people are diagnosed with UTI worldwide, charging the global economy over 6 billion US dollars. The rapid increase in the number of bacteria in the urinary tract is the cause of urinary tract infections (Mollick S *et al*., 2016 & Majumder *et al*, 2022). Many predisposing factors are involved like extremes of age, female gender, pregnancy, instrumentation, urinary tract infection, neurologic dysfunction, renal disease, and expression of A, B, and H blood group oligosaccharides on the surface of epithelial cells. Many virulence factors for urinary pathogens are situated in the human body which encourages adherence to the mucosal surface. This adherence in the mucosal surface ultimately results in subsequent infection (Majumder *et al*, 2022).

Bacteria are responsible as a primary organism that cause UTIs. Gram-negative bacteria cause (80-85%), and gram-positive bacteria cause (10-15%) of the infection (Thangavelu S *et al*, 2022). The major etiological agents of UTIs are *Escherichia coli* and other Enterobacteriaceae. The residual agents are coupled with various organisms, including the Gram-positive bacteria *Staphylococcus saprophyticus*, *E. faecalis, S. agalactiae, S. pyrogens, S. aureus, B. subtilis*, and *Enterococcus* are usually prevalent which is also resistant to antibiotics (Mollick S *et al*, 2016 & Thangavelu S *et al*, 2022).

Females are particularly prone to this disease than males owing to their physical structure specifically their urinary tract structure (Majumder *et al*, 2022 & Naik TB *et al*, 2018). Females are more vulnerable to disease because of their shorter and wider urethra and are easily affected by microorganisms. The traditional rise of bacteria from the fecal flora via the urethra to the bladder and kidney is the main cause of infection (Majumder *et al*, 2022 & Pandey B *et al*, 2020). The disease is more frequent between the ages between 16 and 35. Around 10% of women are infected during their lifetime. It was found that one woman out of five develops a UTI. The re-emergence of the infection is common with almost half of the people will have a secondary infection within a year. Approximately 20% of women who are infected with UTI will have a second attack and more or less 30% of those will experience yet another incident of UTI (Naik TB *et al*, 2018).

Antibiotics have been used successfully for the treatment of bacterial infections in recent years. The appearance of antimicrobial resistance all over the world owing to the abuse of antibiotics has become a serious issue to the health of the human body. Recently, it has been found that more or less 700,000 people worldwide die yearly from antimicrobial resistance (AMR) infections and this number would reach 10 million by 2050 according to the report (Gharavi MJ *et al*, 2021). Although, the death number of UTIs is relatively low because of antimicrobial therapy. But the inappropriate use of antimicrobial agents against UTI can cause risk or future problems associated with MDR. Overuse of antibiotics and unavailability of new drugs for the treatment of the patient has driven the spread of antimicrobial resistance and multidrug resistance (MDR) among urinary pathogenic bacteria like methicillin-resistant *S. aureus* (MRSA), extended-spectrum ß-lactamase (ESBL) *E. coli*, vancomycin-resistant *S. aureus*, and *enterococcus spp*. Gram-positive cocci showed remarkable antibiotic resistance ability in the last decade. Antimicrobial resistance pattern is different in different regions among bacteria producing UTIs. The resistance ability of gram-positive cocci is higher in clinical isolates than in community isolates. The scarcity of information about antimicrobial resistance profiles of gram-positive cocci from clinical isolates has led to the worldwide growing infection rates (Alemayehu T *et al*, 2019 & Hussain T *et al*, 2021)**.** The resistance mechanism of gram-positive and gram-negative bacteria against antibiotics is different. UTI is a very complex disease that is very difficult to treat due to the abundance of resistant uropathogenic bacteria (Gessese YA *et al*, 2017).

Better treatment of patients with UTIs mainly depends on the proper identification of the organisms and the determination of an appropriate antibiotic agent. Adequate knowledge of the etiological agent and their susceptibilities to accessible drugs is of great value of importance. This will lead to proper and effective treatment and ultimately prevent the development of drug resistance (Majumder *et al*, 2022 & Hussain T *et al*, 2021). In Bangladesh, it is sorry to say that it is lacking. In other Asian countries, UTI has been high in pediatrics. The percentage of pediatric patients in Nepal were 15.88% and 57% at different period (Ullah H *et al*, 2022). In India, it was 48%. In Nigeria, the occurrence of UTI among adolescents and children was 11.9% (Ullah H *et al*, 2022). Increased resistance to urinary pathogens has been reported in the eastern region of Bangladesh, India, and Nepal (Addis T *et al*, 2021). The importance of gram-positive cocci has revealed its significance in UTIs according to the previous study (Hussain T *et al*, 2021). Gram-positive bacteria are often found in vulnerable groups such as pregnant women and the elderly as etiologic agents (Naik TB *et al*, 2018 & Urmi UL *et al*, 2019). In Bangladesh, the number of highly valued constructive studies regarding gram-positive bacteria's etiology is low (Urmi UL *et al*, 2019). This study aimed to determine the rate of gram-positive bacterial infections among both genders and examine the patterns of antimicrobial susceptibility of gram-positive bacteria, and also to re-confirm the phenotypic properties of clinically significant isolates of Gram-positive bacteria along with their antimicrobial drug susceptibility pattern. This study will help physicians to choose the proper pragmatic treatment to treat the gram-positive bacteria causing UTI infection.

## 2. Materials and methods

### 2.1 Collection of Samples:

Samples were collected from November 2021 to April 2022. Samples were collected from a highly reputable diagnostic center in Narayanganj, Dhaka, Bangladesh. A total of 1308 urine samples were collected from human patients of different ages and gender who are suspected of having Urinary Tract Infection (UTI). Sample processing and transportation were maintained as per WHO guidelines. All experiments were performed in the Department of Microbiology, Stamford University Bangladesh.

**2.2 Samples processing:** 4 ml of clean midstream urine of each patient were collected in a sterile tube and immediately transferred to the laboratory for investigation. Streaking technique was followed to get the proper pathogenic isolate on various differential and selective media and incubated the media plates for 24 hours at 37°C (Miller JM & Holmes HT, 1999).

**2.3 Isolation and confirmation of Gram-positive bacteria:**

Pure-culture of Gram-positive bacteria and Gram-negative bacteria was isolated and maintained using various differential and selective media like MacConkey agar, blood agar, chocolate agar, salmonella shigella (SS) agar, mannitol salt agar (MSA) media. From 1308 total samples only 121 samples showed gram positive result due to their growth on selective media and those isolates were selected for further analysis (Addis T *et al*, 2021; Urmi UL *et al*, 2019 & Said A *et al*, 2021).

**2.4 Microscopic analysis:**

Microscopic analysis of the isolates was done through bacterial size, shape, and staining properties. Initial identification of selected isolates was performed by gram staining procedure, followed by different biochemical tests. The cultural and morphological characteristics of selected isolates were identified according to standard microbiological protocols (Addis T *et al*., 2021; Mehboob M *et al.*, 2021 & Collee JG *et al*., 1996).

**2.5 Biochemical test for the confirmatory identification:**

All isolated bacteria were identified by standard laboratory biochemical tests according to the methods. The biochemical tests for *S. aureus*, *Streptococcus* spp., *Enterococcus* spp., and beta-hemolytic *Streptococcus* spp. were the indole test, MR-VP test, catalase test, oxidase test, urease test, beta hemolysis test, coagulase test, citrate utilization test, H2S production test as well as mannitol fermentation test (Addis T *et al.,* 2021; Mehboob M *et al.*, 2021 & Collee JG *et al*, 1996).

**2.6 Determination of antimicrobial susceptibility by disk diffusion method**:

In the current study, disc diffusion susceptibility test was performed on Mueller-Hinton agar (MHA; Merck, Germany) based on Clinical and Laboratory Standards Institute documents to determine the susceptibility of UTIs bacteria (Mihankhah A *et al.*, 2017).

Pure culture of GPB (total 121) isolated from urine samples were selected for assaying antimicrobial susceptibility patterns against a different group of antibiotics such as Amikacin (30$\mu$g), Amoxyclav (30$\mu$g), Azithromycin (30$\mu$g). Cefixime (5$\mu$g), Ceftriaxone (30$\mu$g),

Cefuroxime (30μg), Cephradine (30μg), Ciprofloxacin (5μg), Cloxacillin (30μg), Cotrimoxazole (30μg), Doxycycline (30μg), Erythromycin (30μg), Fusidic acid (30μg), Gentamicin (30μg), Meropenem (30μg), Teicoplanin (30μg), Tetracycline (30μg), Vancomycin (30μg), Linezolid (30μg), Tigecycline (30μg), Clindamycin (30μg) and Levofloxacin (30μg).

**Ethical statement**

In this study neither humans/ samples or animals /samples were use.

**3. Result and discussions:**

Urinary tract infections are one of the most common diseases encountered in healthcare settings, causing significant complications and occurring in newborns to the elderly (Mehboob M *et al.*, 2021 and Collee JG *et al.*, 1996 & Mihankhah A *et al.*, 2017). The prevalence of infection varies with age, sex, and certain predispositions. As UTI is a common community-acquired disease in Bangladesh, a significant proportion of the population is infected with the disease each year. This study was designed to review only the antimicrobial susceptibility of Gram-positive UTI pathogens.

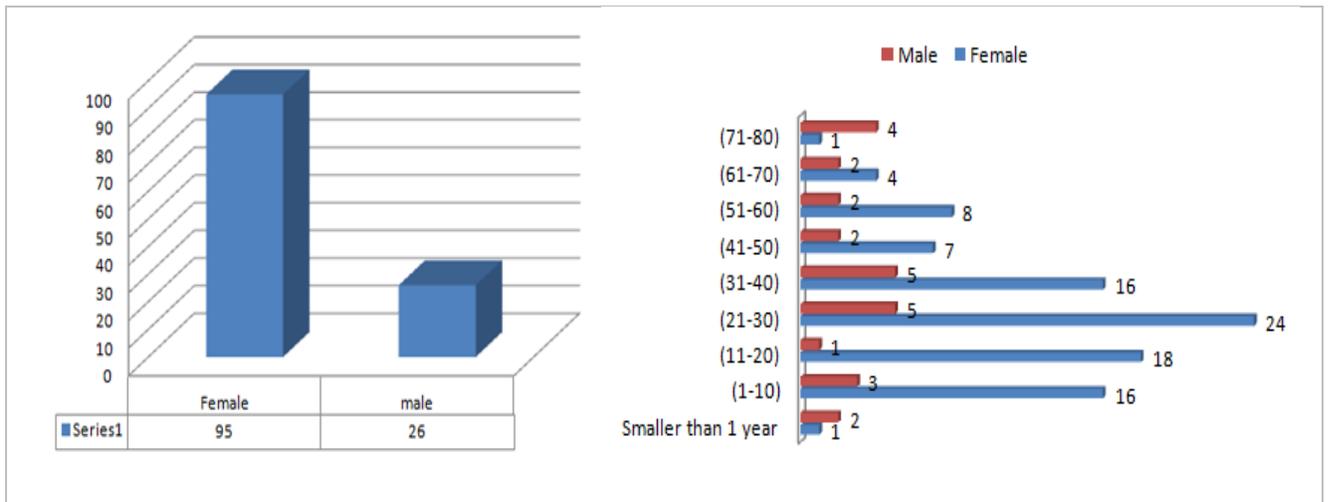

Figure 1. Distribution of (**a**) gender and (**b**) different age groups among respondents.

In this study, a total of 121 isolates were suspected to be suffering from GPB infections of male and female gender with different ages from 1 to 80. We found that female patients are likely to be more susceptible to GPB infection than male patients. Among the isolates tested 95 (78.51%) of them were obtained from female patients and about 26 (21.48%) were obtained from male patients [Fig:1.a].

Females are more UTI prone to their shorter length of Urethra compare to males which allows bacteria to travel less & attach to the urinary tract (Majumder *et al.*, 2022 & Nik TB *et al.*, 2018). Their more Sensitive Skin makes bacteria which can cause irritation & penetration. Specific Types of Contraception pills which help to avoid pregnancy. During menopause women get easily vulnerable to various bacteria and lowering estrogen levels causes susceptibility to bacteria, Pregnancy can make the female body prone to get UTI (Pandey B *et al.*, 2020). In this study, [Fig:1.b] the highest frequency of UTIs (n=24) occurred in females in the age category of 21 to 30. Between (11- 20) years of age group also found higher chances to get UTI occurs in females (Naik TB *et al.*, 2018).

| Isolates | Number |
| --- | --- |
| Total urine sample | 1308 |
| Negative results | 932 |
| Gram-negative bacteria | 255 |
| Gram-positive bacteria | 121 |

Table 1. Distribution based on organism's growth among respondents

| Isolated bacteria | Percentages |
| --- | --- |
| *Enterococcus spp.* | 33.05% |
| *Beta hemolytic streptococci* | 19.00% |
| *Streptococcus spp.* | 20.66% |
| *Staphylococcus aureus* | 27.27% |

Table 2. Microbial isolation of different types of Gram-positive bacteria.

Among 1308 samples almost 932 (71.25%) showed negative result means the patients are not infected with UTI whereas 376 (28.74%) samples showed positive findings. Among which the number of isolation of Gram negative bacteria (GNB) are 255 (19.49%) and Gram positive bacteria (GPB) isolation number was 121(9.25%) respectively [Tab:1].

Though UTIs are mainly caused by gram-negative bacteria, doctors frequently prescribed medicines which is most effective against gram-negative bacteria and also responsible for increasing microbial resistance. Recently, conducted studies show that UTIs can be caused by gram-positive bacteria (Fores-Mireles AL *et al.*, 2015).

But there are a few other gram-positive bacteria like *S. saprophyticus, S. agalactiae, S. pyrogens, S. aureus,* and *B. subtilis* that are usually prevalent, and also cause UTIs (Mollick S *et al.*, 2016). In this study, four types of gram-positive bacteria (*Staphylococcus aureus, Beta-hemolytic streptococci, Enterococcus,* and *Streptococcus* spp.) were found.

In most medical diagnosis centers, the concern for diagnosing GPB is very limited because of the abundance of UTIs caused by GNB (Flores- Mireles AL *et al.*, 2015 & Mollick S *et al.*, 2016). Table 2 shows only the percentage of GPB among all samples respectively *Enterococcus spp.* (33.05%) *Staphylococcus aureus* 27.27%, *Streptococcus* 20.66%, *Beta-hemolytic streptococci (*19.00%) [Fig:2]. Though most of the studies are conducted on GNB & for these reasons doctors mostly prescribe the drugs which work on GNB. But the current findings & the research studies showed the Gram-positive bacteria causing UTI rate was also increasing (Girma A *et al.*, 2022). These findings were consistent with the previous study (Flores- Mireles AL, 2015). Here *Enterococcus* is the highest number among the organisms. (Naik TB *et al.*, 2018 & Said A *et al.*, 2021**)** then *Staphylococcus aureus*, Streptococcus and *Beta-hemolytic streptococci*. These were in accordance with other research conducted in Bangladesh & India (Urmi L *et al.*, 2019).

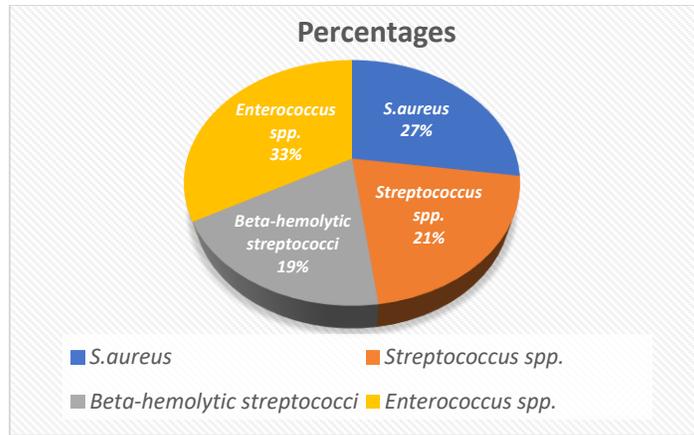

Figure 2. Growth percentage of Gram-positive bacteria among samples

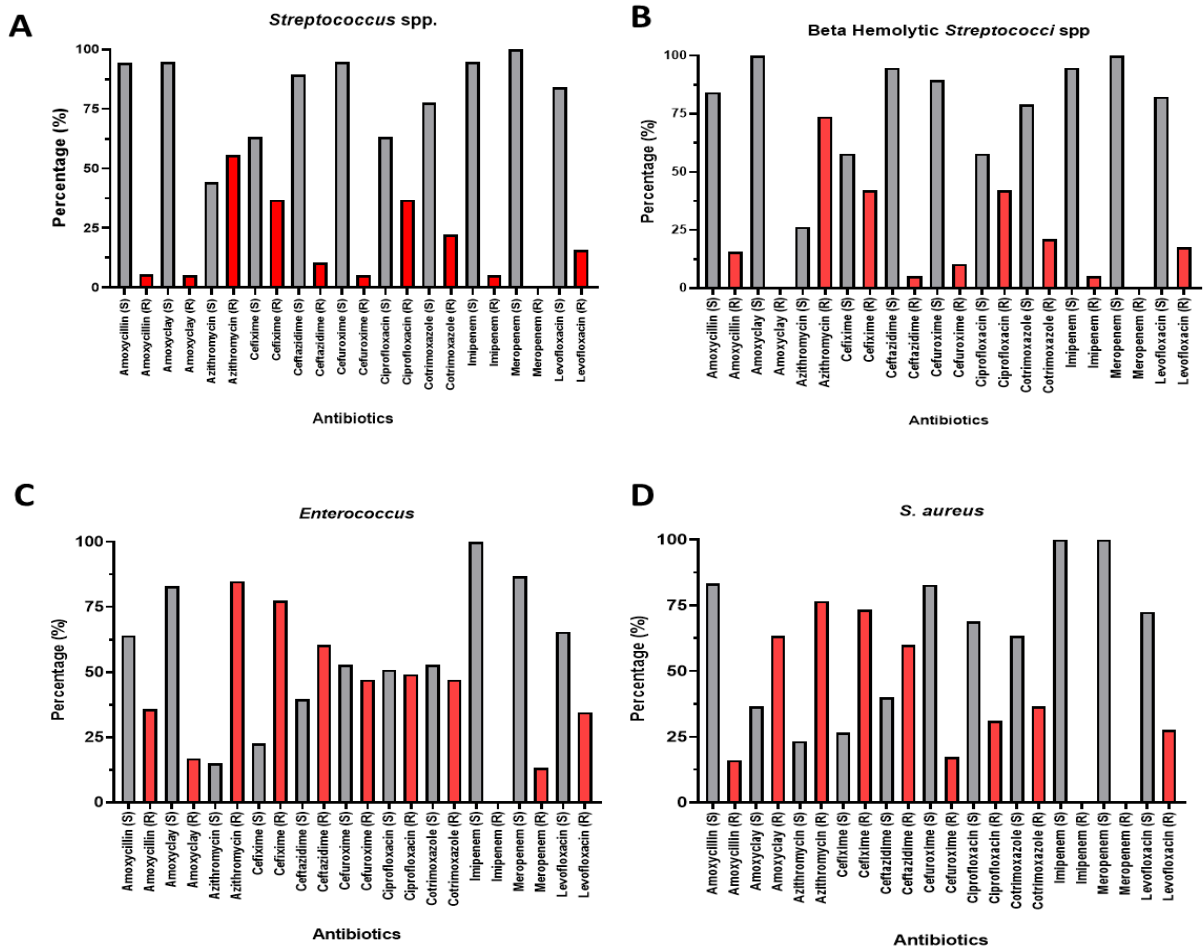

Figure 3. Antibiotic resistant pattern of A. *Streptococcus* spp., B. Beta haemolytic *Streptococcus* spp, C. *Enterococcus* spp., and D. *S. aureus*

Most organisms were resistant to cefixime and azithromycin. In contrast to *Enterococcus spp.*, which exhibited 15% resistance to that antibiotic, Meropenem shown no resistance trend against *Streptococcus spp., S. aureus,* and *beta-hemolytic streptococci. Beta-hemolytic streptococci* and *Streptococcus spp.* demonstrated 5.26% resistance to imipenem, while the other two isolates exhibited no resistance pattern. Amixiclay demonstrated the best 83.33% resistance against *S. aureus* when compared to *Beta-hemolytic streptococci* did not exhibit any antibiotic resistance. Gram-positive bacteria were found in a Iranian investigation revealed they were highly resistant to ampicillin, tetracycline, and erythromycin (92.31%), and highly susceptible to nitrofurantoin and vancomycin (92.3%). However, research conducted in our nation has revealed that a high level of -lactam resistance is a sign of the presence of extended-spectrum -lactamase (ESBL)-producing microbes. Four studies found methicillin-resistant *Staphylococcus aureus* (MRSA). Vancomycin susceptibility of enterococci was reported in three trials, with a median susceptibility of 100%. Penicillin has shown significant sensitivity against Streptococcus pneumoniae (MR 4%) (Khoshbakht R *et al.*, 2013 & Iftekhar Ahmed *et al.*, 2019).

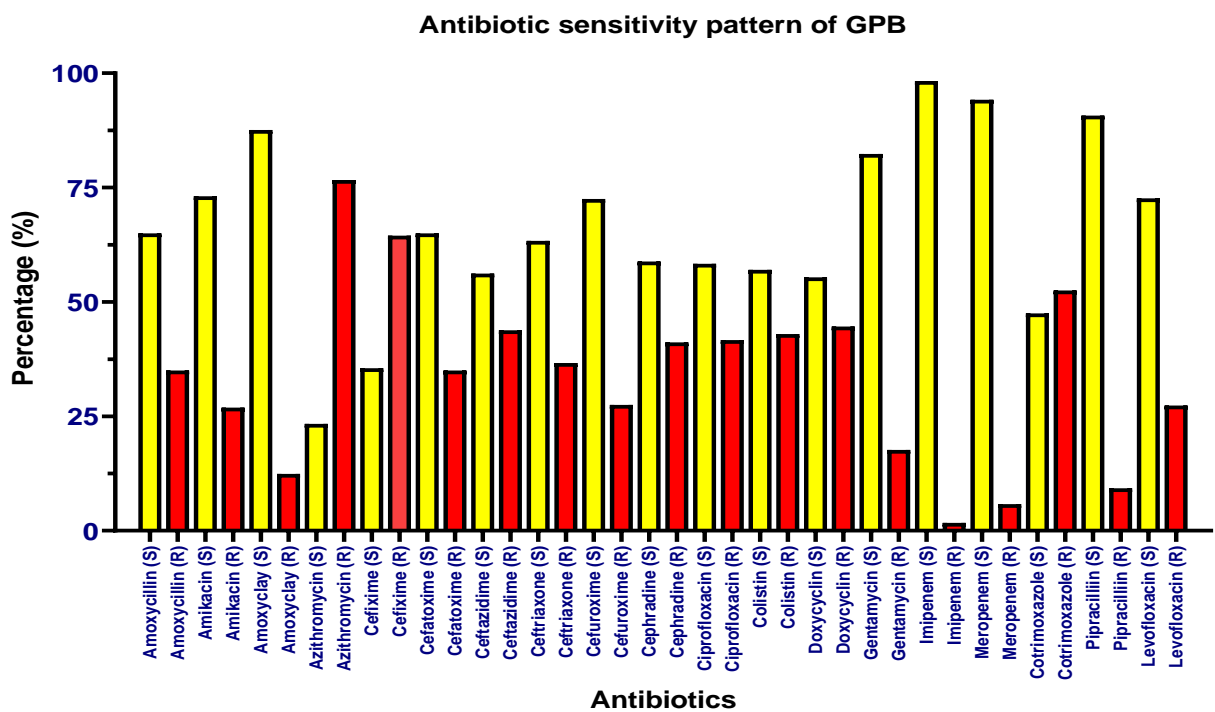

Figure 4. Sensitivity & resistance pattern of GPB isolates against several antibiotics.

Here is the graph which shows overall antibiotic susceptibility pattern of the isolated gram positive bacteria which are S*treptococcus* (15.702%) and *Staphylococcus aureus* (24.79%); *Beta-hemolytic streptococcus* (15.70%), and 43.80% by *Enterococcus* respectively. Generally, the highest susceptibilities of Gram-positive pathogens to antimicrobials were seen towards imipenem by GPB 98% [Fig:3]. Moreover, GPB was found to be highly sensitive to meropenem (94%), piperacillin (90%), amoxiclav (87%), and gentamicin (82%) (Thangavelu S *et al*, 2022; Mehboob M 2021). On the opposite, the lowest susceptibilities of gram-positive pathogens to antimicrobials were seen by azithromycin (23%). Also, sensitivity against Ciprofloxacin (58%), levofloxacin (72%), Ceftriaxone (63%), ceftazidime (56%), %), cephradine (58%), cotrimoxazole (47%), cefuroxime (72%), amikacin (73%), doxycycline (55%), cefixime (35%), cefotaxime (65%), colistin (57%), amoxicillin (65%) [Fig:4].

## 5. Conclusion

UTI is a serious public health issue. The susceptibility pattern of antimicrobials against GPB has outlined the uncontrolled use of antibiotics throughout this study. If it is untreated the burden of UTI which causes drug resistance in general people, especially among the female reproductive age group can bring great consequences. Early diagnosis, proper medication by the experts such as doctors or health professionals, and need some effort to control the misuse of antibiotics and procurement will assist in limiting the increasing rate of antibiotic resistance in pathogenic microorganisms. Also developing further complications will be reduced like the suffering of the patient in a hospital, economic loss, being prone to secondary infection, and infertility if proper treatment is ensured. Apart from this finding, individual assessment of genetic predisposition to UTI, and tailored treatment can also be helpful in the study of genetic variation which may influence drug resistance. It would be of great interest if genetic variation and alteration, metabolism pathway of patients with UTI are possible to study.

## Acknowledgment

Authors would like to thank Stamford University Bangladesh for providing laboratory facilities, technical assistance and financial aid.

## Conflict of Interest

The authors declare no conflict of interest.